\pdfoutput=1

\documentclass[11pt]{article}
\usepackage{multirow, makecell}
\usepackage{tabularx}
\usepackage{times}
\usepackage{latexsym}
\usepackage{booktabs,siunitx,multirow}
\usepackage{enumitem}
\usepackage{xcolor}
\usepackage[most]{tcolorbox}
\usepackage{amsmath}
\usepackage[T1]{fontenc}
\usepackage{graphicx} 
\usepackage{subcaption}
\usepackage{booktabs}
\usepackage{todonotes}
\usepackage{enumitem}
\usepackage{tikz}
\usetikzlibrary{shapes,arrows,positioning}
\usepackage{hyperref}  
\usepackage{nameref}   
\definecolor{lightheadingcolor}{RGB}{70,130,180}
\definecolor{boxcolor}{RGB}{240,248,255}

\usepackage[utf8]{inputenc}
\usepackage{xspace}
\usepackage{microtype}
\usepackage[usestackEOL]{stackengine}
\newcolumntype{P}[1]{>{\centering\arraybackslash}p{#1}}

\usepackage[utf8]{inputenc} 
\usepackage[T1]{fontenc}    
\usepackage{hyperref}       
\usepackage{url}            
\usepackage{booktabs}       
\usepackage{amsfonts}       
\usepackage{nicefrac}       
\usepackage{microtype}      

\usepackage[preprint]{acl}

\usepackage{times}
\usepackage{latexsym}

\usepackage[T1]{fontenc}

\usepackage[utf8]{inputenc}

\usepackage{microtype}

\usepackage{inconsolata}

\usepackage{graphicx}

\title{Multi-Hop Question Answering: When Can Humans Help, and Where do They Struggle?}

\author{Jinyan Su$^1$, 
Claire Cardie$^1$, Jennifer Healey$^2$\\
  $^1$ Cornell University, 
  $^2$ Adobe Research,\\
  \texttt{\{js3673, ctc9\}@cornell.edu, jehealey@adobe.com}
  }
\begin{document}
\maketitle
\begin{abstract}
Multi-hop question answering is challenging task for both large language models (LLMs) and humans, as it requires recognizing when multi-hop reasoning is needed, followed by reading comprehension, logical reasoning, and knowledge integration. To better understand how humans might collaborate effectively with AI, we evaluate the performance of crowd workers on these individual reasoning subtasks. We find that while humans excel at knowledge integration (97\% accuracy), they often fail to recognize when a question requires multi-hop reasoning (67\% accuracy). Participants perform reasonably well on both single-hop and multi-hop QA (84\% and 80\% accuracy, respectively), but frequently make semantic mistakes—for example, answering “when” an event happened when the question asked “where.” These findings highlight the importance of designing AI systems that complement human strengths while compensating for common weaknesses.
\end{abstract}

\section{Introduction}
\label{sec:intro}
With the rapid development of large language models (LLMs), retrieval-augmented generation (RAG) has emerged as a popular approach for open-domain question answering \cite{lewis2020retrieval, shi2023replug}. By combining parametric knowledge stored within pre-trained LLMs and non-parametric knowledge retrieved from external sources, RAG improves answer accuracy, and reduces hallucinations. 

However, multi-hop question answering—where answering a query requires integrating information from multiple sources—remains a significant challenge \cite{trivedi2022interleaving} for both humans and LLMs. Multi-hop reasoning combines reading comprehension, logical reasoning, and knowledge integration, making it substantially more complex than single-hop QA. Handling multi-hop queries within RAG systems often requires more sophisticated retrieval mechanisms, such as iterative retrieval \cite{asai2023self, trivedi2022interleaving}, or more powerful models with enhanced reasoning and planning capabilities \cite{jiang2024retrieve, su2025fast}. While these approaches can improve answer quality, they come at the cost of increased computational expenses and longer response time \cite{jeong2024adaptive, su2025fast}.

To balance cost and response accuracy, many existing systems employ adaptive retrieval strategies, where simple queries are handled with lightweight retrieval or no retrieval at all, whereas complex queries invoke more expensive, multi-step retrieval and reasoning pipelines \cite{su2025fast, jeong2024adaptive}. While this improves efficiency, such adaptive approaches lack flexibility and user control. In real-world scenarios, user preferences can vary: some may prefer fully automated responses, while others may prioritize speed and accuracy and be willing to assist in the process—by decomposing complex queries, performing manual retrieval, or integrating multiple retrieved results themselves. Such human-AI collaboration would require people to be able to understand the LLM's reasoning process, including: recognizing query complexity, decomposing complex queries, and synthesizing retrieved answers into a final response.


This work aims to assess people's ability collaborate with AI across different aspects of the reasoning process.  We investigate:
\begin{itemize}
\item \textbf{Q1.} Can people effectively recognize multi-hop complexity?

\item \textbf{Q2.} Which reasoning steps: reading comprehension, logical reasoning, or knowledge integration are most challenging for people? 
\end{itemize}

\begin{figure*}[tbh]
    \centering
   \includegraphics[width=0.9\linewidth]{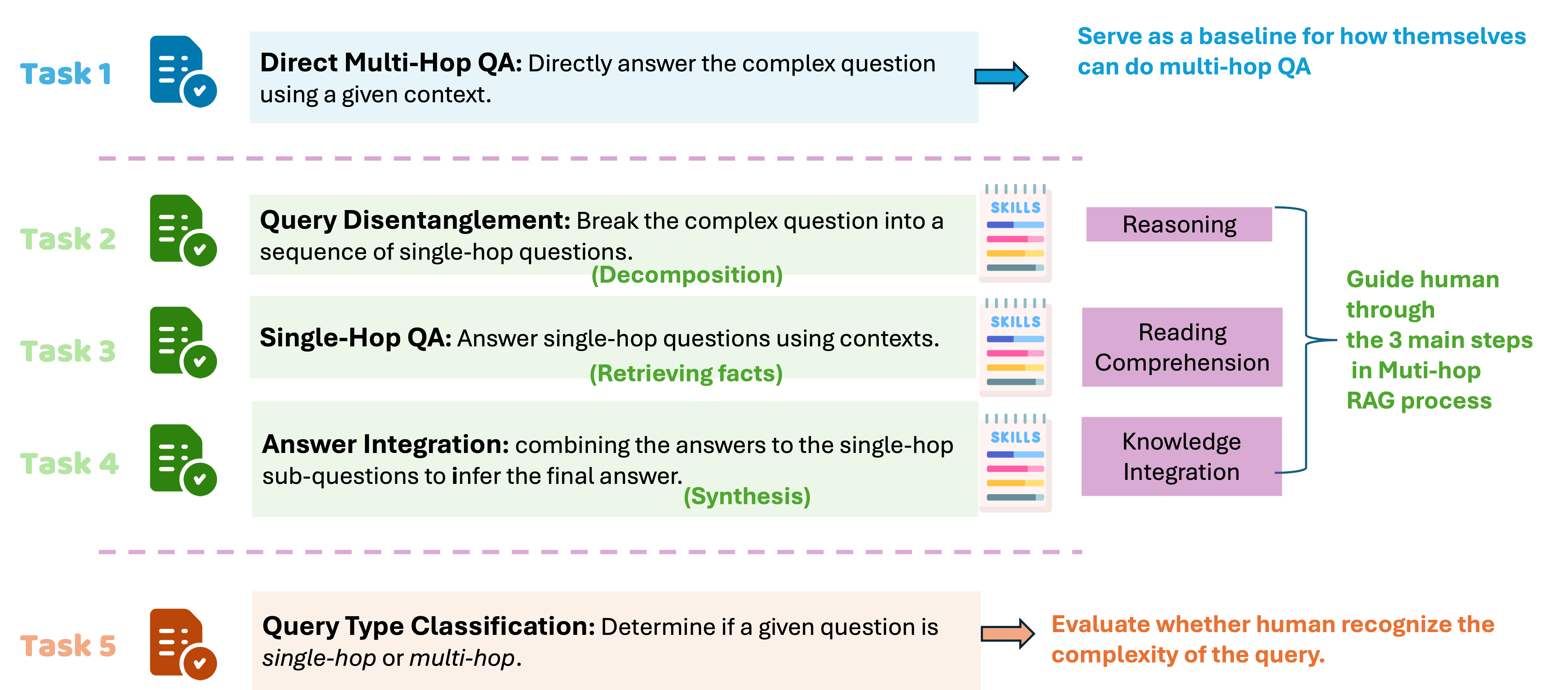}
    \caption{\textbf{Overview of the five human evaluation tasks designed to analyze how humans perform in multi-hop retrieval-augmented generation (RAG)}. Task 1 serves as a baseline, asking participants to directly answer complex multi-hop questions. Tasks 2–4 decompose the RAG pipeline into core steps—query decomposition, single-hop answering, and answer integration—requiring capabilities in reasoning, reading comprehension, and knowledge synthesis. After guiding the participants through the complete multi-hop RAG stages, Task 5 evaluates whether participants can recognize the complexity of a query (single-hop vs. multi-hop), a prerequisite for adaptive retrieval strategies. These tasks help identify which steps are most challenging for humans and where human intervention is most valuable, informing the design of collaborative RAG systems.}
    \label{fig: task}
\end{figure*}

We designed a multi-task user study to answer these questions. The study comprised of five sequential tasks, 
illustrated in Figure \ref{fig: task}, that introduce participants to the concept of complex (multi-hop) versus simple (single-hop) questions and evaluate their ability to effectively perform various sub-tasks.  
Our findings aim to support the development of user-controllable QA workflows that allow users to dynamically trade off response time, reasoning effort, and answer quality. Our contributions include:
\begin{itemize}
\item We provide the first fine-grained empirical study on human effort in multi-hop QA that identifies the relative difficulty of query type identification,  reading comprehension, question decomposition and answer synthesis. 
\item We find that humans can directly answer multi-hop questions (Task 1) with 80.2\% accuracy and perform single-hop QA (Task 3) with 84.1\% accuracy. Participants were best at answer integration (Task 4), reaching 97.3\% accuracy, but not as good at query decomposition (Task 2), at 78.2\% accuracy. 
\item We find that query complexity (Task 5), was the most difficult achieving only 67.9\% accuracy for a two choice answer. This suggests that humans are unreliable at selecting the appropriate RAG strategy, motivating the need for automated, adaptive RAG systems.
\end{itemize}
Our findings have implications for designing more efficient and collaborative QA systems, where users and AI work together to improve reasoning speed, cost efficiency, and answer accuracy. 

\section{Related Work}
\textbf{Multi-Hop QA}
Multi-hop question answering (QA) requires both comprehension and reasoning across multiple pieces of evidence \cite{trivedi2022interleaving}. 
Early benchmarks revealed that automated systems struggled with these tasks, with model performance lagging significantly behind that of humans \cite{yang2018hotpotqa, ho2020constructing, li2024meqa}. However, prior work typically evaluated human performance using a small, non-representative sample—often limited to just a few participants (e.g., three graduate students). This narrow sampling fails to reflect the broader user base of retrieval-augmented generation (RAG) systems or the general population. Furthermore, existing studies did not systematically assess human performance on the constituent subtasks of multi-hop QA, nor did they evaluate whether humans could accurately distinguish between single-hop and multi-hop questions.

\section{Human Study Task Design}
\paragraph{Dataset}
We base our evaluation on the 2WikiMultiHopQA dataset \cite{ho2020constructing}, a multi-hop question answering benchmark constructed from Wikipedia and Wikidata. Each question is designed to require reasoning over multiple documents, and the dataset includes structured evidence paths that explicitly define the multi-hop reasoning required. Compared to earlier datasets like HotpotQA \cite{yang2018hotpotqa}, 2WikiMultiHopQA more rigorously enforces multi-hop dependencies and has been widely used in recent work on retrieval-augmented generation (RAG) with large language models \cite{su2025fast, trivedi2022interleaving, jeong2024adaptive}. This makes it an ideal foundation for analyzing how humans perform in multi-hop QA and where they can contribute in hybrid systems.

\subsection{Task Overview}\label{sec: task overview}
We design five tasks to evaluate human performance across the stages on multi-hop QA. The task order is fixed to ensure the sub-tasks are introduced in a particular sequence. Each task, described in detail in Appendix \ref{sec:surveyinstructions}, discusses complex (multi-hop) and simple (single-hop) queries. The first task (Task 1) introduces the overall problem of multi-hop reasoning. This is followed by tasks (Task 2-4) that introduce reasoning sub-problems: fact retrieval, question decomposition and answer integration.  
The tasks are designed to  familiarize participants with the concept of multi-hop retrieval. 
Finally, participates are asked to classify queries as being either simple (single-hop) or complex (multi-hop) in Task 5. This final task assesses whether humans can reliably recognize query complexity, a critical prerequisite for user-directed retrieval decisions. 
\paragraph{Task 1: Direct Multi-Hop QA(DMH)}
In the first task, participants are given a full multi-hop query along with its context paragraphs that provide the facts necessary to answer the query. This simulates a standard QA setup, where no intermediate steps are shown or required. This serves as a baseline for human multi-hop QA and allows us to compare the performance on the overall problem to performance on the various subtasks.
\paragraph{Tasks 2–4: Decomposing and Supporting Multi-Hop Reasoning}
To better understand where humans struggle or excel, we break the multi-hop QA process into three core components. These stages reflect the internal structure of many pipeline-style RAG systems and allow us to isolate distinct cognitive operations: structuring the reasoning process (Task 2), retrieving facts (Task 3), and synthesizing intermediate answers (Task 4). By evaluating each step independently, we can identify which part of the reasoning chain imposes the greatest challenge for humans, and where human input might be most valuable in collaborative QA systems.

\vspace*{0.1in}
\noindent
\textbf{\textit{Task 2: Query Decomposition(QD).}} Participants are given a complex multi-hop question and asked to rewrite it as a sequence of simpler, single-hop sub-questions. Each sub-question should isolate a specific reasoning step that contributes to answering the original query. This task assesses participants' ability to structure multi-step reasoning into modular components—a skill analogous to planning in decomposition-based QA systems. Accurate decomposition is critical for multi-hop pipelines, as errors in this step can propagate through retrieval and synthesis.

\vspace*{0.1in}
\noindent
\textbf{\textit{Task 3: Single-Hop QA(SH).}} In this task, participants are provided with single-hop questions that have been pre-generated from multi-hop queries (i.e., they do not perform decomposition themselves). Each question is accompanied by the original context (10–20 paragraphs) from the corresponding multi-hop instance. Participants are asked to answer each single-hop question independently using the provided context.
This task isolates factual retrieval: it evaluates whether humans can correctly extract or infer answers when the reasoning step has already been made explicit. It reflects the role of retrievers or extractors in RAG systems, which are often responsible for returning evidence when given a well-formed query.

\vspace*{0.1in}
\noindent
\textbf{\textit{Task 4: Answer Integration(AInt).}} In this task, participants are given the original multi-hop question, a set of corresponding single-hop sub-questions, and the answers to those sub-questions. Using only this structured information—without access to the original context—they are asked to produce a final answer to the multi-hop query. This task isolates the final step of multi-hop reasoning: integrating intermediate facts to resolve the overall question. Unlike previous tasks, participants are not responsible for retrieval or decomposition here; their sole focus is on combining sub-answers corresponding to each step of the reasoning chain. This mirrors the synthesis stage in RAG systems, where multiple retrieved facts or intermediate answers must be logically combined to produce the final response.

\paragraph{Task 5: Query Type Classification (QTC)}
In the final task, participants are shown queries and asked to classify each as either single-hop or multi-hop. After completing Tasks 1 through 4, they are expected to have developed a clear sense of what distinguishes simple and complex queries. This task evaluates whether humans can recognize query complexity—an important prerequisite for enabling user-controlled retrieval strategies in hybrid QA systems.
If humans can make this distinction reliably, it suggests they could choose between lightweight and multi-step RAG pipelines themselves. If not, it supports the need for automated adaptive-RAG systems that dynamically adjust the retrieval strategy based on question complexity.

\subsection{Human Evaluation Setup}
We implemented our tasks using a survey generated with Quartics and recruited 40 participants through Prolific.  Participants were required to be fluent in English and were paid a rate of \$12 USD per hour for the task which was estimated at 40 minutes duration. The order of the tasks was intentionally fixed to introduce a progression of tasks designed to educate the participant toward understanding the difference between "simple" (single hop) and complex (multi-hop) questions.  Within each of the tasks, the presentation of questions was randomized.  The survey included two attention check questions to ensure that the users were human and were reading the instructions.  All users passed the attention check criteria.  One user, whose overall score was less than two standard deviations lower than the mean score, was excluded for contributing answers that did not make sense.    

\section{Results}
Table~\ref{tab:survey_results} summarizes the accuracy of the participants in performance each tasks described in Section \ref{sec: task overview}.  As previously reported in the ~\ref{sec:intro}, humans excel at AInt (97.3\% accuracy)and struggle most with Query type classification (67.9\% accuracy for a two choice question).  The average time per query per task across all participants was: DMH 97.2s, QD 20.2s, SH 138.8s, AInt 9.2s, and QTC 11.0s.  

\begin{table}[ht]
    \centering
    \caption{Accuracy statistics for each crowdworker task: direct multi-hop QA (DMH), single-hop QA (SH), query disentanglement (QD), answer integration (AInt), and question type classification (QTC).}
    \label{tab:survey_results}
    \renewcommand{\arraystretch}{1.2}
    \begin{tabular}{l|c|c|c|c|c}
        \toprule
        \textbf{} & \textbf{DMH} & \textbf{SH} & \textbf{QD} & \textbf{AInt} & \textbf{QTC} \\
        \midrule
        Mean & 80.2 & 84.1 & 78.2 & 97.3 & 67.9 \\
        Min  & 70.1 & 76.9 & 64.1 & 94.1 & 51.3 \\
        Max  & 91.9 & 100  & 89.7 & 100  & 82.1 \\
        \bottomrule
    \end{tabular}
\end{table}

\noindent For DMH we initially considered six queries, but eliminated one because the correct answer (Austria) was not included in its paragraphs\footnote{question\_id: 350ef4460bde11eba7f7acde48001122}. 
For both DMH and SH the most common errors were answering the wrong question (e.g. answering "who" for a question that asked "when") and confusing similarly named people. For QD, the most common error was failing to include an integration question, especially when the integration would be obvious for a person (e.g when asking for a comparison, failing to ask a question that compares the answers). For AInt, incorrect answers were rare and did not follow a pattern. For QTC we found the accuracy score surprisingly low, given the participants prior exposure to multiple examples of simple and complex questions.  Participants were slightly better at identifying the three Simple questions (77.8\% accuracy) than the four complex questions (65.3\% accuracy).   

\section{Discussion}
This study was designed to assess people’s ability collaborate with AI across different aspects of the multi-hop query reasoning process. We found that while people excelled at answer integration and performed well on both single and multi-hop question answering, they missed steps in question decomposition and were often unable to recognize when a question required multiple steps. 
In collaborative systems, it is likely that AI will have to identify when a question is complex and requires multiple steps.  From there, our findings indicate that people would likely be able to decompose the identified complex question into simple questions and integrate the answers. When people themselves are generating the questions it is likely that they will not be confused about the desired fact (e.g. "where" vs. "when") but if humans are being asked to act as evaluators, it is more likely that LLMs could help double check human answers and catch such errors.  Designing more
effective and collaborative QA systems, where people and AI work together, requires a nuanced understanding of the strengths and weaknesses of each to produce the best results.
\section*{Limitations}
While our study offers valuable insights into human strengths and weaknesses in multi-hop question answering, the participant pool—40 English-speaking crowd workers recruited via Prolific—may not fully capture the diversity of potential end-users in terms of language background, education, or reasoning strategies. Nevertheless, our sample reflects a realistic segment of the non-expert user population likely to interact with retrieval-augmented QA systems in practice. Future work can build on this foundation by exploring more diverse populations and use contexts.
\section*{Ethical Statement}
All participants were recruited through Prolific and provided informed consent before beginning the task. Participants were compensated at a fair hourly rate (\$12/hour), exceeding the platform’s minimum wage requirement. The tasks involved non-sensitive, publicly available text data, and no personal or identifying information was collected. The study posed minimal risk to participants.
\bibliography{custom}

\newpage
\appendix

\onecolumn
\section{Appendix: Survey Instructions}
\label{sec:surveyinstructions}
In this section, we provide the specific instructions that were given to crowd-worker participants. The following page (Box~\ref{box:survey-instructions}) is the overview page we give the participate before starting the survey. 
\begin{tcolorbox}[enhanced,colback=boxcolor,colframe=lightheadingcolor,
                 fonttitle=\bfseries\large,
                 title=Survey Instructions Overview Page,
                 label=box:survey-instructions
                 ]

\subsection*{Welcome to our Survey!}

\textbf{Overview:} This survey consists of 5 tasks, each containing 6 questions. While some tasks are more extensive than others, we estimate the entire survey can be completed in approximately \textbf{20 minutes}.

\subsection*{Survey Focus}
This survey examines simple and complex questions:
\begin{itemize}[leftmargin=*]
  \item \textbf{Simple question:} Requires only one fact to answer
  \item \textbf{Complex question:} Requires multiple facts to be discovered
\end{itemize}

\vspace{0.5cm}
\begin{center}
\begin{tabular}{p{6cm}|p{6cm}}
\toprule
\textbf{Example: Complex Question} & \textbf{Example: Simple Questions} \\
\midrule
"Where was the 40th President of the United States born?" & "Who was the 40th President of the United States?" \\
 & "Where was Ronald Reagan born?" \\
\bottomrule
\end{tabular}
\end{center}
\vspace{0.5cm}

\subsection*{Task Types}
Participants will engage in tasks such as:
\begin{itemize}[leftmargin=*]
  \item Answering complex questions given a list of facts
  \item Creating simple questions from complex questions
  \item Distinguishing between simple and complex questions
\end{itemize}

\textbf{Note:} All tasks are timed. If you need a break, please take it on the instruction page before each task.

\subsection*{Research Purpose}
We are investigating the difficulty of these tasks when participants are provided with a list of facts. Task completion time serves as our primary measure of difficulty.

\subsection*{Logistics}
 You will enter your Prolific ID at the beginning of the survey. Your ID will be saved by the survey system. Following completion, you will be redirected back to Prolific. Both your answers and Prolific ID will be recorded. The survey contains two attention check questions. Responses will be verified within 48 hours. Payment will be cleared if you pass at least one of two attention checks

\textbf{To continue:} Please enter your Prolific ID on the next screen.
\end{tcolorbox}

\vspace{1cm}
Then, before staring each task, the participate are given instructions and specific examples for each task. 
\vspace{1em}

\begin{tcolorbox}[
    enhanced,
    colback=blue!5!white, 
    colframe=blue!80!black, 
    colbacktitle=blue!70!black,
    title={\Large\textbf{Task 1: Direct Multi-Hop QA}},
    fonttitle=\color{white}\bfseries,
    sharp corners=south,
    boxrule=1pt,
    drop shadow={black!50!white},
    label=box:task1
]

\begin{center}
\textcolor{blue!80!black}{\LARGE\textbf{Instructions}}
\end{center}
\noindent\rule{\linewidth}{1pt}

\begin{itemize}
    \item You will be given a Question and a set of Facts. Use the Facts to answer the question.
    \item You must connect information from multiple facts to find the answer.
    \item The Question appears at the beginning and end of the Facts for convenience.
    \item This is a timed task. Please work efficiently.
\end{itemize}

\begin{tcolorbox}[
    enhanced,
    colback=blue!2!white,
    colframe=blue!40!black,
    title={\textbf{Example}},
    fonttitle=\color{white},
    colbacktitle=blue!40!black
]

\textbf{Question:} Where was the 40th President of the United States born?

\vspace{0.5em}
\textbf{Facts:}
\begin{itemize}[leftmargin=2em, itemsep=0pt]
    \item William Jefferson Clinton (né Blythe; born August 19, 1946) is an American politician and lawyer who served as the 42nd president...
    \item Ronald Wilson Reagan was an American politician and actor who served as the 40th president of the United States from 1981 to 1989...
    \item George Herbert Walker Bush (June 12, 1924 – November 30, 2018) was the 41st president...
    \item Born in Illinois, President Reagan graduated from Eureka College in 1932...
    \item James Earl Carter Jr. (October 1, 1924 – December 29, 2024) was an American politician...
    \item Born and raised in Arkansas, President Clinton graduated from Georgetown University...
\end{itemize}

\vspace{0.5em}
\textbf{Question:} Where was the 40th President of the United States born?

\vspace{0.5em}
\begin{tcolorbox}[
    colback=blue!3!white,
    colframe=green!50!black,
    title={\textbf{Your Answer}},
    fonttitle=\color{white},
    colbacktitle=blue!30!black,
    boxrule=0.5pt
]
\textbf{Answer:} \colorbox{green!10}{\textit{Illinois}}

\vspace{0.3em}
\textbf{Reasoning:} First, we need to identify that Ronald Reagan was the 40th President (fact 2), then located his birthplace in fact 4 which states he was born in Illinois.
\end{tcolorbox}
\end{tcolorbox}
\end{tcolorbox}
\begin{tcolorbox}[
    enhanced,
    colback=blue!5!white, 
    colframe=blue!80!black, 
    colbacktitle=blue!70!black,
    title={\Large\textbf{Task 2: Query Disentanglement}},
    fonttitle=\color{white}\bfseries,
    sharp corners=south,
    boxrule=1pt,
    drop shadow={black!50!white},
    label=box:task2
]

\begin{center}
\textcolor{blue!80!black}{\LARGE\textbf{Instructions}}
\end{center}
\noindent\rule{\linewidth}{1pt}

\begin{itemize}
    \item Given a Complex Question, break it down into at least two Simple Questions, each retrieving a necessary fact.
    \item You may refer to previous answers as A1, A2, etc.
\end{itemize}

\begin{tcolorbox}[
    enhanced,
    colback=blue!2!white,
    colframe=blue!40!black,
    title={\textbf{Example 1}},
    fonttitle=\color{white},
    colbacktitle=blue!40!black
]

\textbf{Complex Question:} Where was the wife of the President of the United States after Jimmy Carter born?

\vspace{0.5em}
\begin{tcolorbox}[
    colback=blue!3!white,
    colframe=blue!30!black,
    title={\textbf{Your Answer: }},
    fonttitle=\color{white},
    colbacktitle=blue!30!black,
    boxrule=0.5pt
]
\begin{enumerate}[leftmargin=*, label=\textbf{Simple Question \arabic*:}]
    \item \colorbox{green!10}{Who was the President of the United States after Jimmy Carter?}
    \item \colorbox{green!10}{Who was the wife of A1?}
    \item \colorbox{green!10}{Where was A2 born?}
    \item \textless not needed, leave blank\textgreater
\end{enumerate}
\end{tcolorbox}
\end{tcolorbox}

\begin{tcolorbox}[
    enhanced,
    colback=blue!2!white,
    colframe=blue!40!black,
    title={\textbf{Example 2}},
    fonttitle=\color{white},
    colbacktitle=blue!40!black
]

\textbf{Complex Question:} Which film, "Goonies" or "Pretty in Pink", was released first?

\vspace{0.5em}
\begin{tcolorbox}[
    colback=blue!3!white,
    colframe=blue!30!black,
    title={\textbf{Your Answer}},
    fonttitle=\color{white},
    colbacktitle=blue!30!black,
    boxrule=0.5pt
]
\begin{enumerate}[leftmargin=*, label=\textbf{Your Answer \arabic*:}]
    \item \colorbox{green!10}{When was "Goonies" released?}
    \item \colorbox{green!10}{When was "Pretty in Pink" released?}
    \item \colorbox{green!10}{Which year is earlier A1 or A2?}
    \item \colorbox{green!10}{Which film was released in A3?}
\end{enumerate}
\end{tcolorbox}
\end{tcolorbox}
\end{tcolorbox}
\vspace{1em}

\begin{tcolorbox}[
    enhanced,
    colback=blue!5!white, 
    colframe=blue!80!black, 
    colbacktitle=blue!70!black,
    title={\Large\textbf{Task 3: Single-Hop QA}},
    fonttitle=\color{white}\bfseries,
    sharp corners=south,
    boxrule=1pt,
    drop shadow={black!50!white},
    label=box:task3
]

\begin{center}
\textcolor{blue!80!black}{\LARGE\textbf{Instructions}}
\end{center}
\noindent\rule{\linewidth}{1pt}

\begin{itemize}
    \item You will be given facts and three questions (two Simple questions , one Complex question).
    \item The Complex Question can be answered from the answers to the Simple Questions.
\end{itemize}

\begin{tcolorbox}[
    enhanced,
    colback=blue!2!white,
    colframe=blue!40!black,
    title={\textbf{Example}},
    fonttitle=\color{white},
    colbacktitle=blue!40!black
]

\begin{tcolorbox}[
    colback=blue!3!white,
    colframe=blue!30!black,
    title={\textbf{Questions}},
    fonttitle=\color{white},
    colbacktitle=blue!30!black,
    boxrule=0.5pt
]
\begin{enumerate}[leftmargin=*, label=\textbf{Question \arabic*:}]
    \item (Simple) When was "Goonies" released? \hfill (Answer:  \colorbox{green!10}{1985})
    \item (Simple) When was "Pretty in Pink" released? \hfill (Answer: \colorbox{green!10}{1986})
    \item (Complex) Which was released first: "Goonies" or "Pretty in Pink"? \hfill (Answer: \colorbox{green!10}{Goonies})
\end{enumerate}
\end{tcolorbox}

\vspace{0.5em}
\begin{tcolorbox}[
    colback=blue!3!white,
    colframe=blue!30!black,
    title={\textbf{Facts}},
    fonttitle=\color{white},
    colbacktitle=blue!30!black,
    boxrule=0.5pt
]
\begin{itemize}
    \item On June 7, 1985, "The Goonies" was released in theaters...
    \item "Pretty in Pink" was released February 28, 1986...
\end{itemize}
\end{tcolorbox}
\end{tcolorbox}
\end{tcolorbox}

\vspace{1em}

\begin{tcolorbox}[
    enhanced,
    colback=blue!5!white, 
    colframe=blue!80!black, 
    colbacktitle=blue!70!black,
    title={\Large\textbf{Task 4: Answer Integration}},
    fonttitle=\color{white}\bfseries,
    sharp corners=south,
    boxrule=1pt,
    drop shadow={black!50!white},
    label=box:task4
]

\begin{center}
\textcolor{blue!80!black}{\LARGE\textbf{Instructions}}
\end{center}
\noindent\rule{\linewidth}{1pt}

\begin{itemize}
    \item Answer a Complex Question by integrating answers from its Simple Questions.
\end{itemize}

\begin{tcolorbox}[
    enhanced,
    colback=blue!2!white,
    colframe=blue!40!black,
    title={\textbf{Example}},
    fonttitle=\color{white},
    colbacktitle=blue!40!black
]

\textbf{Complex Question:} Who was born first, Albert Einstein or Thomas Edison?

\vspace{0.5em}
\begin{tcolorbox}[
    colback=blue!3!white,
    colframe=blue!30!black,
    title={\textbf{Simple Questions with Answers}},
    fonttitle=\color{white},
    colbacktitle=blue!30!black,
    boxrule=0.5pt
]
\begin{enumerate}[leftmargin=*, label=\textbf{Simple Question \arabic*:}]
    \item When was Albert Einstein born? \hfill (A1: March 14, 1879)
    \item When was Thomas Edison born? \hfill (A2: February 11, 1847)
\end{enumerate}
\end{tcolorbox}

\vspace{0.5em}
\begin{tcolorbox}[
    colback=blue!3!white,
    colframe=green!50!black,
    title={\textbf{Answer to Complex Question}},
    fonttitle=\color{white},
    colbacktitle=blue!30!black,
    boxrule=0.5pt
]
\textbf{Answer:} \colorbox{green!10}{Thomas Edison}
\end{tcolorbox}
\end{tcolorbox}
\end{tcolorbox}

\vspace{1em}

\begin{tcolorbox}[
    enhanced,
    colback=blue!5!white, 
    colframe=blue!80!black, 
    colbacktitle=blue!70!black,
    title={\Large\textbf{Task 5: Query Type Classification}},
    fonttitle=\color{white}\bfseries,
    sharp corners=south,
    boxrule=1pt,
    drop shadow={black!50!white},
    label=box:task5
]
\begin{center}
\textcolor{blue!80!black}{\LARGE\textbf{Instructions}}
\end{center}
\noindent\rule{\linewidth}{1pt}

\begin{itemize}
    \item Classify each query as Simple (single fact needed) or Complex (multiple facts needed).
\end{itemize}

\begin{tcolorbox}[
    enhanced,
    colback=blue!2!white,
    colframe=blue!40!black,
    title={\textbf{Example 1}},
    fonttitle=\color{white},
    colbacktitle=blue!40!black
]

\textbf{Query:} When was the music group the Beatles originally formed?

\vspace{0.5em}
\begin{tcolorbox}[
    colback=blue!3!white,
    colframe=blue!30!black,
    title={\textbf{Your Answer}},
    fonttitle=\color{white},
    colbacktitle=blue!30!black,
    boxrule=0.5pt
]
\colorbox{green!10}{\textbf{Simple Question}}

\textit{(This query requires only one fact: when the Beatles formed)}
\end{tcolorbox}
\end{tcolorbox}

\begin{tcolorbox}[
    enhanced,
    colback=blue!2!white,
    colframe=blue!40!black,
    title={\textbf{Example 2}},
    fonttitle=\color{white},
    colbacktitle=blue!40!black
]

\textbf{Query:} When was the music group with Paul McCartney, Pete Best, George Harrison and John Lennon originally formed?

\vspace{0.5em}
\begin{tcolorbox}[
    colback=blue!3!white,
    colframe=blue!30!black,
    title={\textbf{Your Answer}},
    fonttitle=\color{white},
    colbacktitle=blue!30!black,
    boxrule=0.5pt
]
\colorbox{green!10}{\textbf{Complex Question}}

\textit{(This query requires multiple facts: identifying the band from its members, then finding when it formed)}
\end{tcolorbox}
\end{tcolorbox}
\end{tcolorbox}

\end{document}